\documentclass[twocolumn,english]{revtex4}
\usepackage{mathptmx}
\usepackage[T1]{fontenc}
\usepackage[latin9]{inputenc}
\usepackage{amsmath}
\usepackage{amssymb}
\usepackage{graphicx}

\makeatletter

\newcommand{\noun}[1]{\textsc{#1}}

\@ifundefined{textcolor}{}
{%
 \definecolor{BLACK}{gray}{0}
 \definecolor{WHITE}{gray}{1}
 \definecolor{RED}{rgb}{1,0,0}
 \definecolor{GREEN}{rgb}{0,1,0}
 \definecolor{BLUE}{rgb}{0,0,1}
 \definecolor{CYAN}{cmyk}{1,0,0,0}
 \definecolor{MAGENTA}{cmyk}{0,1,0,0}
 \definecolor{YELLOW}{cmyk}{0,0,1,0}
}

\usepackage{empheq}

\usepackage{babel}

\usepackage{babel}

\usepackage{babel}

\makeatother

\usepackage{babel}
\begin{document}

\title{{\huge{}{{}{Topologically protected excitons in porphyrin thin
films}}}}

\author{Joel Yuen-Zhou$^{1}$, Semion K. Saikin$^{1,2}$, Norman Yao$^{3}$,
and Alán Aspuru-Guzik$^{1,2}$}

\affiliation{$^{1}$Center for Excitonics, Research Laboratory of Electronics,
Massachusetts Institute of Technology, Cambridge, MA, USA.}
\email{joelyuen@mit.edu}

\affiliation{$^{2}$Department of Chemistry and Chemical Biology, Harvard University,
Cambridge, MA, USA.}

\affiliation{$^{3}$Department of Physics, Harvard University, Cambridge, MA,
USA.}
\begin{abstract}
The control of exciton transport in organic materials is of fundamental
importance for the development of efficient light-harvesting systems.
This transport is easily deteriorated by traps in the disordered energy
landscape. Here, we propose and analyze a system that supports topological
Frenkel exciton edge states. Backscattering of these chiral Frenkel
excitons is prohibited by symmetry, ensuring that the transport properties
of such a system are robust against disorder. To implement our idea,
we propose a two-dimensional periodic array of tilted porphyrins interacting
with a\emph{ }homogenous magnetic field. This field serves to break
time-reversal symmetry and results in lattice fluxes that that mimic
the Aharonov-Bohm phase acquired by electrons. Our proposal is the
first blueprint for realizing topological phases of matter in molecular
aggregates and suggests a paradigm for engineering novel excitonic
materials.
\end{abstract}
\maketitle
Upon interactions with light, molecules are promoted to excited states,
typically referred to as molecular excitons \cite{davydovbook} (hereafter
referred to as excitons interchangeably). The efficient transport
of energy across a molecular array via such excitons is one of the
main goals in the design of solar cell devices \cite{gregg}. This
owes to the fact that excitons mediate energy transfer between incoming
photons and the electrical current generated upon excited-state dissociation
\cite{scholesrumbles,semion_excitonics}. In this context, tremendous
efforts have been focused on organic photovoltaic materials, which
have advantages in terms of production cost, chemical versatility,
and enhanced absorption properties. However, an important drawback
in such organic materials is the presence of a large amount of disorder
\cite{bredasopv,brabec}. Indeed, static disorder arises from structural
imperfections in the molecular aggregate which yield local perturbations
to the on-site energies as well as to the couplings between molecules.
Typically, such disorder induces Anderson localization of the single-particle
exciton eigenstates \cite{fidder:7880}, which significantly reduces
transport. While such localizing effects can be partially compensated
by the addition of vibrations which help untrap the exciton \cite{mohseni,plenio,cao},
a generic solution remains a challenge.

Our approach to this challenge draws upon ideas from the field of
disordered electronic systems ---in particular, from the phenomenology
broadly termed as ``quantum Hall effects'' (QHEs) \cite{qhebook}.
A hallmark of such quantum hall systems is that they exhibit delocalized
current-carrying chiral edge modes. Specifically, the breaking of
time-reversal symmetry (TRS) in these systems ensures that there are
no counter-propagating modes to backscatter into \cite{halperin}.
Elegant extensions of these ideas include photonic setups \cite{wangsoljacic,hafezi,rechtsman,hafezi_2,khanikaev}
and topological insulators (TI) --- materials that preserve TRS but
whose edge modes are related to strong spin-orbit coupling \cite{hasankane,qizhang,bernevigbook,kong}.
We note that organometallic TIs have recently been suggested by Liu
and coworkers \cite{doi:10.1021/nl401147u,liu_natcom,PhysRevLett.110.106804,PhysRevLett.110.196801},
paving the way towards a wider and possibly cheaper group of materials
that may exhibit these exotic phenomena.

Since QHEs have been posed in the context of electrons and photons,
it is natural to inquire whether their excitonic analog exists. The
present article answers this question positively, by explicitly constructing
a minimal model of a Frenkel exciton porphyrin lattice which supports
topologically protected edge states when it interacts with a magnetic
field. Since this effort is already challenging by itself, we limit
ourselves to cryogenic temperatures, and therefore, disregard effects
of vibrational dephasing of excitons, which we shall study elsewhere.
As far as we are aware, this is curiously the first work that addresses
the joint effects of both \emph{magnetic fields} and \emph{coherence}
in molecular exciton transport. Furthermore, this article is also
the first example of topological phases in molecular excitons, and
therefore, offers a novel approach to the design of a new generation
of materials for more efficient energy harvesting and transport.

\section*{Description of the model}

\begin{figure}
\begin{centering}
\includegraphics[scale=0.35]{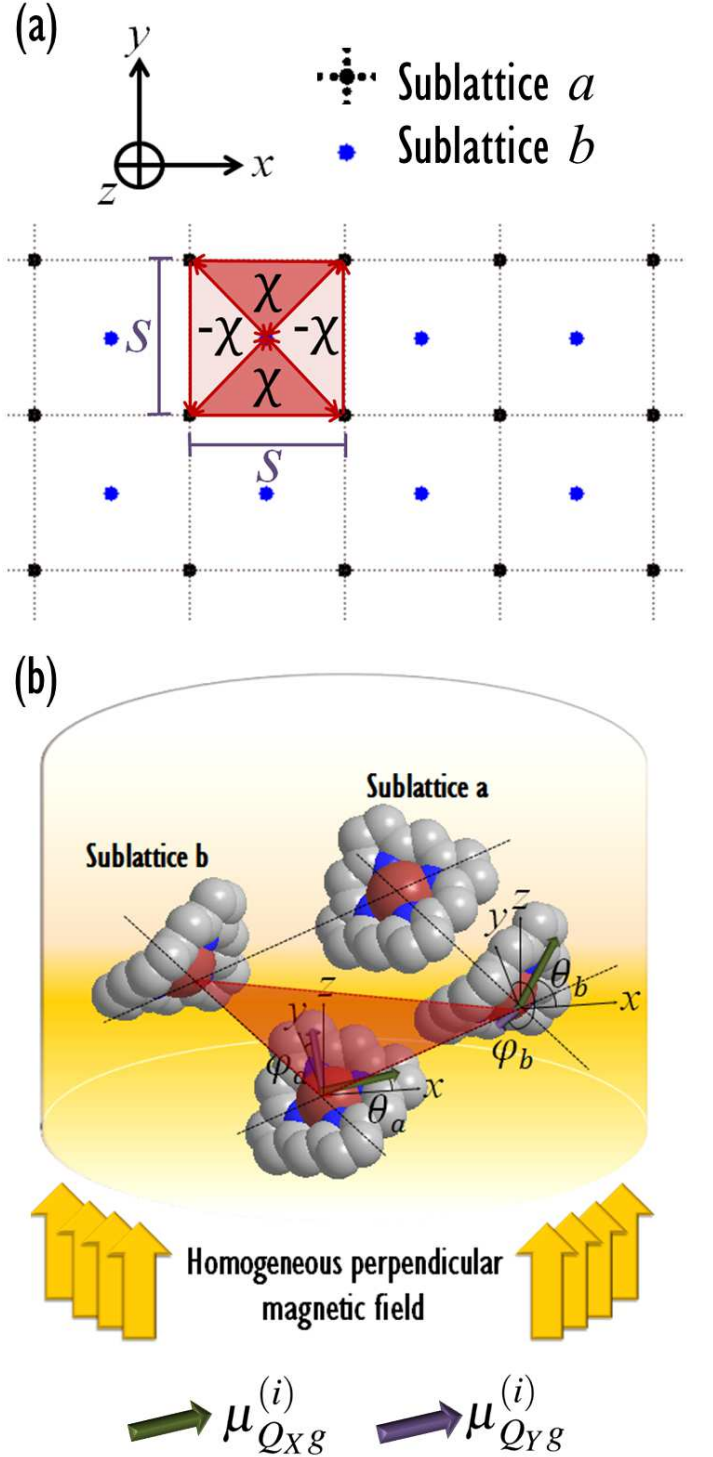}
\par\end{centering}

\protect\protect\caption{\emph{Porphyrin lattice under a uniform magnetic field.} (a) The square
lattice consists of porphyrins arranged into two sublattices $a$
(black) and $b$ (blue). A possible unit cell is a square of dimensions
$s\times s$. We highlight the fluxes (Berry phases) $\pm\chi$ that
the exciton obtains upon counterclockwise circulation about each of
minimal three porphyrin loops. This pattern of fluxes is reminiscent
to Haldane's model, where electrons are placed in a lattice under
an inhomogeneous magnetic field with no effective magnetic field per
unit cell, and yet, manifest nontrivial topological properties. (b)
Magnification of the upper corner of the unit cell in (a). The actual
setup uses a homogeneous magnetic field. Sublattice $a$ differs from
sublattice $b$ in the orientation angles $\theta_{i},\varphi_{i}$
($i=a,b$) of the porphyrins with respect to the $x$ and $y$ axes.
Each porphyrin consists of two transition dipoles $\boldsymbol{\mu}_{Q_{X}g}^{(i)}$
and $\boldsymbol{\mu}_{Q_{Y}g}^{(i)}$ in the plane of the molecules.
The anisotropic character of dipolar interactions together with the
magnetic field makes the exciton hoppings along the northeast and
the northwest directions different enough to obtain a nonzero flux
$\chi$. \label{fig:lattice}}
\end{figure}

\emph{Description of the lattice.---} Our setup consists of a two-dimensional
periodic array of unsubstituted metalloporphyrins (hereafter referred
to just as porphyrins), molecules with $D_{4h}$ symmetry that maintain
their planarity due to their metal centers \cite{porphyrins,multiporphyrin},
and which are well known compounds in photovoltaic applications \cite{porphyrin_photovoltaic,porphyrin_photovoltaic_2,porphyrin_photovoltaic_3}.
These porphyrins are arranged in a square lattice in the $\boldsymbol{xy}$
plane with a unit cell of area $s\times s$ (see Fig. \ref{fig:lattice}a),
where $s\sim\frac{1}{2}-2\,\mbox{nm}$. The lattice consists of two
sublattices $a$ and $b$, where the porphyrins are tilted out of
the $\boldsymbol{x}\boldsymbol{y}$ plane in ways that depend on two
angles per sublattice, $\theta_{i}$ and $\varphi_{i}$ ($i=a,b$),
respectively. This two-dimensional lattice can in principle be realized
by self-assembly techniques exploiting an already crystalline substrate
\cite{jacs_porphyrins,CPHC:CPHC200700494,molecular_model,tilt,salvan,Feyer201464},
which in our case, shall be chosen to avoid exciton quenching processes
(an insulating material fullfills this condition \cite{small_maier,Scarfato2008677}).

Using a Cartesian vector notation in the ``lab'' or array frame
throughout the article, the $a$ sites are located at positions $\boldsymbol{n}s\equiv(n_{x},n_{y},0)s$
for $n_{x},n_{y}$ integers, whereas the $b$ sites are at $(\boldsymbol{n}+\frac{1}{2})s=(n_{x}+\frac{1}{2},n_{y}+\frac{1}{2},0)s$.
We shall be concerned with the three lowest electronic states in each
molecule, namely, its ground state $|g^{(i)}\rangle$, and its degenerate
Q-band absorbing in the visible spectrum ($\omega_{Q}\sim17350\,\mbox{cm}^{-1}$
\cite{Malley1968320}), consisting of the orthogonal states $|Q_{X}^{(i)}\rangle$
and $|Q_{Y}^{(i)}\rangle$ (we use capital labels for Cartesian coordinates
for the molecular frame of each sublattice).

Due to the degeneracy of the Q-band, the states $|Q_{X}^{(i)}\rangle$
and $|Q_{Y}^{(i)}\rangle$ can be arbitrarily defined as long as their
transition dipole moments with respect to $|g^{(i)}\rangle$ constitute
an orthogonal set of vectors of equal magnitude $d$ spanning the
plane of each porphyrin. We denote the transition dipole operator
for a porphyrin in sublattice $i$ by $\boldsymbol{\mu}^{(i)}=\boldsymbol{\mu}_{Q_{X}g}^{(i)}|Q_{X}^{(i)}\rangle\langle g^{(i)}|+\boldsymbol{\mu}_{Q_{Y}g}^{(i)}|Q_{Y}^{(i)}\rangle\langle g^{(i)}|+\mbox{c.c.}$,
where $\boldsymbol{\mu}_{Q_{X}g}^{(i)}=\boldsymbol{\mu}_{gQ_{X}}^{(i)}$
and $\boldsymbol{\mu}_{Q_{Y}g}^{(i)}=\boldsymbol{\mu}_{gQ_{Y}}^{(i)}$
are chosen such that $\boldsymbol{\mu}_{Q_{X}g}^{(i)}$ has zero projection
along the $y$ axis, and $\boldsymbol{\mu}_{Q_{Y}g}^{(i)}$ is orthogonal
to it,
\begin{eqnarray*}
\boldsymbol{\mu}_{Q_{X}g}^{(i)} & = & d(\mbox{cos}\theta_{i},0,\mbox{sin}\theta_{i}),\\
\boldsymbol{\mu}_{Q_{Y}g}^{(i)} & = & d(-\mbox{sin}\varphi_{i}\mbox{sin}\theta_{i},\mbox{cos}\varphi_{i},\mbox{sin}\varphi_{i}\mbox{cos}\theta_{i}),
\end{eqnarray*}
where $d\sim2-8\,\mbox{D}$ depending on the chemical environment
of the porphyrins \cite{vauthey}. These vectors define molecular
frames for each sublattice, with Cartesian unit vectors $\boldsymbol{X}_{i}=\frac{\boldsymbol{\mu}_{Q_{X}g}^{(i)}}{d}$,
$\boldsymbol{Y}_{i}=\frac{\boldsymbol{\mu}_{Q_{Y}g}^{(i)}}{d}$, and
$\boldsymbol{Z}_{i}=\boldsymbol{X}_{i}\times\boldsymbol{Y}_{i}$.
Also, in general $(\theta_{a},\varphi_{a})\neq(\theta_{b},\varphi_{b})$,
so the tilting angles distinguish the sublattices.

\emph{Interaction with a magnetic field.---} Before dealing with the
dipolar interactions between the different porphyrins, we consider
their Zeeman interaction with a perpendicular and homogeneous magnetic
field $\boldsymbol{B}=(0,0,B)$ (see Fig. \ref{fig:dressing}). This
implicitly assumes a separation of timescales whereby the Zeeman effect
is much stronger than the interchromophoric interaction \cite{canters},
\begin{equation}
H^{(i)}=H_{0}^{(i)}-\mu_{0}\boldsymbol{B}\cdot(\boldsymbol{L}^{(i)}+2\boldsymbol{S}^{(i)}).\label{eq:light_matter}
\end{equation}
Here, $H_{0}^{(i)}=\omega_{Q}(|Q_{X}^{(i)}\rangle\langle Q_{X}^{(i)}|+|Q_{Y}^{(i)}\rangle\langle Q_{Y}^{(i)}|)$
is the bare Hamiltonian of each porphyrin, $\mu_{0}=0.47\,\mbox{cm}^{-1}\mbox{T}^{-1}$
is the Bohr magneton, and $\boldsymbol{L}^{(i)}$ and $\boldsymbol{S}^{(i)}$
are the electronic orbital angular momentum and spin of the $i$th
porphyrin. Each of the three states per molecule is a singlet state
with $\boldsymbol{S}^{(i)}=0$. It is valid to regard the porphyrins
as approximate rings in the $\boldsymbol{X}_{i}\boldsymbol{Y}_{i}$
planes occupied by 18 electrons \cite{Malley1968320,rodriguez:205102,canters}.
This implies that the solutions to Eq. (\ref{eq:light_matter}) are
states with approximately good angular momentum quantum number $L_{Z}^{(i)}=\pm m$
perpendicular to the plane of the molecules at $\boldsymbol{Z}_{i}$
for integer $m$ (here $\hbar=1$). In particular, we have $L_{Z}^{(i)}|g^{(i)}\rangle=0$
and $L_{Z}^{(i)}|Q_{\pm}^{(i)}\rangle\equiv L_{Z}^{(i)}(|Q_{X}^{(i)}\rangle\pm i|Q_{Y}^{(i)}\rangle)/\sqrt{2}=\pm9|Q_{\pm}\rangle$,
yielding $H^{(i)}|g^{(i)}\rangle=0$ and $H^{(i)}|Q_{\pm}^{(i)}\rangle=(\omega_{Q}\mp\Delta_{i})|Q_{\pm}^{(i)}\rangle$,
where half of the Zeeman splitting is given by $\Delta_{i}\equiv9\mu_{0}B_{z}\kappa_{i}$,
and we have used $\kappa_{i}\equiv\boldsymbol{z}\cdot\boldsymbol{Z_{i}}=\mbox{cos}\theta_{i}\mbox{cos}\varphi_{i}$.
That is, under a magnetic field, the degenerate Q-band in each porphyrin
splits into two Zeeman levels $|Q_{\pm}^{(i)}\rangle$ with different
energies. Notice that due to TRS breaking, their coefficients in terms
of the ``bare'' states $|Q_{X}^{(i)}\rangle$ and $|Q_{Y}^{(i)}\rangle$
are in general complex. Although not essential, we simplify the model
by fixing the projection of the magnetic field on both sublattices
to be a constant $|\kappa_{i}|=\kappa\neq0$, yielding a constant
Zeeman splitting throughout $|\Delta_{i}|=\Delta$. There is however,
a possibly different ordering of the $|Q_{\pm}^{(i)}\rangle$ states,
depending on the sign of $B_{z}\kappa_{i}$. With this in mind, each
porphyrin has states of energy $\omega_{L}\equiv\omega_{Q}-\Delta$
and $\omega_{U}\equiv\omega_{Q}+\Delta$, which we call the \emph{lower}
and \emph{upper} energy states $|Q_{L}^{(i)}\rangle$ and $|Q_{U}^{(i)}\rangle$,
and $(L,U)=(+,-)$ if $B_{z}\kappa_{i}>0$ and $(L,U)=(-,+)$ otherwise.
Working under a magnetic field of $|B_{z}|=10\,\mbox{T}$, this splitting
attains a value of $2\Delta\sim84\,\mbox{cm}^{-1}$, which is confirmed
by magnetic dichroism experiments \cite{Malley1968320} (the reference
reports half of the actual splitting due to isotropic averaging).
It is clear from this model that other chromophores with similar electronic
structure, such as metallophthalocyanines \cite{porphyrins}, can
be used instead of metalloporphyrins.

\begin{center}
\begin{figure}
\begin{centering}
\includegraphics[scale=0.25]{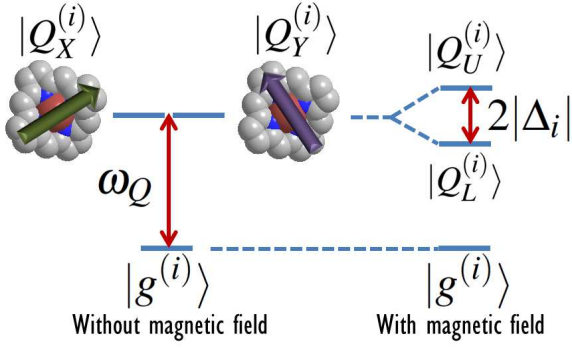}
\par\end{centering}

\protect\protect\caption{\emph{(Metallo-)Porphyrin under a magnetic field.} In the absence
of a magnetic field, a porphyrin can be thought of as three-level
molecules with a ground state $|g^{(i)}\rangle$ and a Q-band formed
by two degenerate excited states $|Q_{X}^{(i)}\rangle$ and $|Q_{Y}^{(i)}\rangle$
at an energy $\omega_{Q}$. A magnetic field $\boldsymbol{B}$ breaks
time-reversal symmetry, as well as the degeneracy via a Zeeman splitting
$2|\Delta_{i}|$, yielding lower $|Q_{L}^{(i)}\rangle$ and upper
$|Q_{U}^{(i)}\rangle$ eigenstates with definite angular momentum
along $\boldsymbol{Z}_{i}$. Here, $i$ denotes different sublattices
$i=a,b$. \label{fig:dressing}}
\end{figure}

\par\end{center}

\emph{Dipolar interactions between Zeeman levels.--- }At this point,
we turn our attention to interactions between porphyrin transitions
across the lattice. We are only interested in weak, ``single-excitation''
effects, so it is convenient to introduce the the global ground state
$|G\rangle\equiv|g\rangle\cdots|g\rangle$ as well as the single-site
excited states $|\boldsymbol{n}_{q}^{(i)}\rangle\equiv|g\rangle\cdots|Q_{q}^{(i)}\rangle\cdots|g\rangle$,
with $q=X,Y,L,U$, where every porphyrin in the lattice is in the
ground state except for the porphyrin in sublatice $i$ with coordinates
$\boldsymbol{n}s$, which is in the $|Q_{q}^{(i)}\rangle$ excited
state \cite{may-kuhn}. Couplings between the bare porphyrin sites
in the lattice are well approximated by the classical real-valued
dipole-dipole interaction \cite{valkunas},
\begin{equation}
\langle\boldsymbol{n}_{q}^{(i)}|\mathcal{J}|\boldsymbol{m}_{r}^{(j)}\rangle\approx\frac{f}{|\boldsymbol{R}_{nm}^{(ij)}|^{3}}\Big(\boldsymbol{\mu}_{Q_{q}g}^{(i)}\cdot\boldsymbol{\mu}_{Q_{r}g}^{(j)}-3(\boldsymbol{\mu}_{Q_{q}g}^{(i)}\cdot\boldsymbol{e}{}_{nm}^{(ij)})(\boldsymbol{\mu}_{Q_{r}g}^{(j)}\cdot\boldsymbol{e}_{nm}^{(ij)})\Big).\label{eq:dipole-dipole}
\end{equation}
Here, $f=5.04\,\mbox{cm}^{-1}(\mbox{nm}^{3}/\mbox{D}^{2})$ (we have
set the index of refraction to 1), $q,r=X,Y$, and $\boldsymbol{R}_{nm}^{(ij)}$
is the position vector pointing from the $\boldsymbol{n}$th porphyrin
of sublattice $i$ to the $\boldsymbol{m}$th porphyrin of sublattice
$j$, and $\boldsymbol{e}_{nm}^{(ij)}=\frac{\boldsymbol{R}_{nm}^{(ij)}}{|\boldsymbol{R}_{nm}^{(ij)}|}$.
To establish an energy scale associated with these interactions, we
define $J\equiv\frac{fd^{2}}{s^{3}}$, which for $d=3\,\mbox{D}$
and $s=2\,\mbox{nm}$ gives $J=5.7\,\mbox{cm}^{-1}$, so that $2\Delta\gg J$,
the Zeeman spitting is much larger than the dipolar couplings, consistent
with our assumptions. Couplings between Zeeman levels follow from
Eq. (\ref{eq:dipole-dipole}) and the appropriate change of basis,

\begin{equation}
\langle\boldsymbol{n}_{u}^{(i)}|\mathcal{J}|\boldsymbol{m}_{v}^{(j)}\rangle=\sum_{q,r=X,Y}\langle Q_{u}^{(i)}|Q_{q}^{(i)}\rangle\langle\boldsymbol{n}_{q}^{(i)}|\mathcal{J}|\boldsymbol{m}_{r}^{(j)}\rangle\langle Q_{r}^{(j)}|Q_{v}^{(j)}\rangle,\label{eq:change_of_basis}
\end{equation}
where $u,v=L,U$. From the energetic considerations above, we need
to include couplings within each band of $|Q_{L}^{(i)}\rangle$ or
$|Q_{U}^{(i)}\rangle$ states, but not between them. Therefore, we
may write a Frenkel exciton Hamiltonian for the total lattice that
reads as $\mathcal{H}=\mathcal{H}_{L}+\mathcal{H}_{U}$. As an illustration,
let us explicitly construct the Hamiltonian $\mathcal{H}_{L}$ for
the lower energy states $|Q_{L}^{(i)}\rangle$, or alternatively $|\boldsymbol{n}_{L}^{(i)}\rangle$,
by introducing the second quantized notation for each sublattice $a_{\boldsymbol{n}}^{\dagger}|G\rangle=|\boldsymbol{n}_{L}^{(a)}\rangle$
and $b_{\boldsymbol{n}}^{\dagger}|G\rangle=|\boldsymbol{n}_{L}^{(b)}\rangle$.
Restricting the interactions to nearest neighbors (NN) and next-nearest
neighbors (NNN), we have the following two-band model,

\begin{eqnarray}
\mathcal{H}_{L} & = & \sum_{n}\Bigg(\omega_{L}(a_{\boldsymbol{n}}^{\dagger}a_{\boldsymbol{n}}+b_{\boldsymbol{n}}^{\dagger}b_{\boldsymbol{n}})\nonumber \\
 &  & +J_{ab,NE}(a_{\boldsymbol{n}+\boldsymbol{NE}}^{\dagger}b_{\boldsymbol{n}}+a_{\boldsymbol{n}-\boldsymbol{NE}}^{\dagger}b_{\boldsymbol{n}})\nonumber \\
 &  & +J_{ab,NW}(a_{\boldsymbol{n}+\boldsymbol{NW}}^{\dagger}b_{\boldsymbol{n}}+a_{\boldsymbol{n}-\boldsymbol{NW}}^{\dagger}b_{\boldsymbol{n}})\nonumber \\
 &  & +J_{aa,E}a_{\boldsymbol{n}+\boldsymbol{E}}^{\dagger}a_{\boldsymbol{n}}+J_{aa,N}a_{\boldsymbol{n}+\boldsymbol{N}}^{\dagger}a_{\boldsymbol{n}}\nonumber \\
 &  & +J_{bb,E}b_{\boldsymbol{n}+\boldsymbol{E}}^{\dagger}b_{\boldsymbol{n}}+J_{bb,N}b_{\boldsymbol{n}+\boldsymbol{N}}^{\dagger}b_{\boldsymbol{n}}\Bigg)+\mbox{c.c.},\label{eq:Hamiltonian_lattice}
\end{eqnarray}
where the couplings are given by $J_{ij,\boldsymbol{V}}\equiv\langle(\boldsymbol{n}+\boldsymbol{V})_{L}^{(i)}|\mathcal{J}|\boldsymbol{n}_{L}^{(j)}\rangle=\langle(\boldsymbol{n}-\boldsymbol{V})_{L}^{(i)}|\mathcal{J}|\boldsymbol{n}_{L}^{(j)}\rangle$,
and the vectors $\boldsymbol{V}$ for the north, east, northeast,
and northwest directions are $\boldsymbol{N}=(0,1,0)$, $\boldsymbol{E}=(1,0,0)$,
$\boldsymbol{NE}=\frac{1}{2}(1,1,0)$, and $\boldsymbol{NW}=\frac{1}{2}(-1,1,0)$.
The analogous Hamiltonian $\mathcal{H}_{U}$ can be similarly constructed
using the states $|Q_{U}^{(i)}\rangle$ or $|\boldsymbol{n}_{U}^{(i)}\rangle$.
It is easy to check that the NNN couplings $J_{ii,\boldsymbol{V}}$
are real-valued, but that the NN couplings $J_{ab,\boldsymbol{V}}$
are complex-valued in general due to the overlaps $\langle Q_{u}^{(i)}|Q_{q}^{(i)}\rangle$
associated with the change of basis (see Eq. (\ref{eq:change_of_basis})).
We note that complex phases in these couplings do not represent physical
observables on their own, as they can be modified via gauge transformations.
Yet, the Berry phases accumulated in closed loops, and in particular
those obtained by encircling the minimal three porphyrins loops are
gauge invariant modulo $2\pi$ and therefore, have observable consequences.
Fig. \ref{fig:lattice} shows that in each unit cell, two of the counterclockwise
loops yield a phase $\chi\equiv\mbox{arg}(J_{ab,NW}^{*}J_{aa,NE}J_{ab,E})=-\mbox{arg}(J_{ab,NW}J_{ab,E})$
and the other two yield the opposite phase $\mbox{arg}(J_{ab,SE}J_{ba,SW}J_{aa,N})=-\chi$.
A peculiar feature of dipolar interactions between the tilted porphyrins
is its anisotropic character which renders $\mbox{arg}(J_{ab,NE})$
different from $\mbox{arg}(J_{ab,NW})$, except for a measure-zero
set of critical orientations, and therefore keeps $\chi$ finite for
every set of tilting angles as long as the magnetic field is on. This
observation is also at the core of recent work on topological phases
in dipolar spins for optical lattices \cite{nyao,nyao2}. Thus, Eq.
(\ref{eq:Hamiltonian_lattice}) has the same structure as the Hamiltonian
for electrons in a lattice under a perpendicular and \emph{inhomogeneous}
magnetic field threading net fluxes $\pm\chi$ across the minimal
loops, but a net zero magnetic flux per unit cell, and therefore,
across the lattice. Let us summarize what we have done so far: we
have constructed a model where Frenkel excitons (quasiparticles with
no net charge) under a homogenous magnetic field behave as if they
were electrons (particles with charge) in an inhomogeneous magnetic
field. Hence, this model is a square lattice version of Haldane's
honeycomb problem, and therefore, expected to exhibit nontrivial topological
properties, in particular, chiral Frenkel exciton edge states which
are robust against disorder \cite{haldane}. This realization is the
main finding of the article.

\section*{Discussion}

Upon fixing the Zeeman splitting $2\Delta=|B_{z}|\kappa$, the unit
cell distance $s$, and transition dipole moment strength $d$, there
is a two-dimensional parameter space which is left to be explored
by varying the tilting angles $\theta_{a}$ and $\theta_{b}$ (the
possible values of $\varphi_{a}$ and $\varphi_{b}$ are fixed by
$\kappa$). These parameters suffice to provide a topological characterization
of $\mathcal{H}_{v}$ via the Chern numbers $\mathcal{C}_{v}$ for
$v=L,U$ (see Supplementary Information for details) \cite{tknn,bernevigbook}.
The physical meaning of $\mathcal{C}_{v}$ is the following: Its sign
denotes the chirality of the edge states (in our convention, positive
for clockwise and negative for counterclockwise); its magnitude is
equal to the (integer) number of edge states per value of quasimomentum
(given open boundary conditions (OBC) along one axis).

\begin{figure}
\begin{centering}
\includegraphics[scale=0.15]{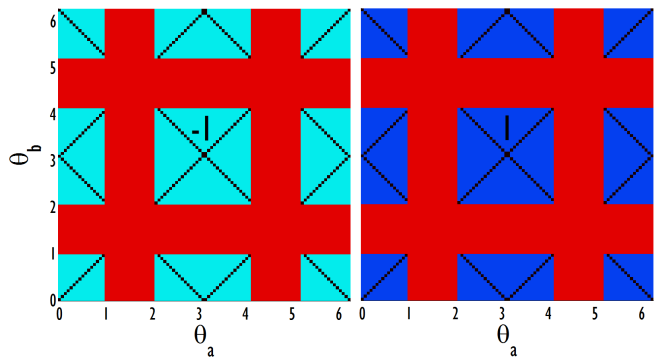}
\par\end{centering}

\protect\protect\caption{\emph{Phase diagram of exciton topological phases.} We fix\emph{ $\kappa=|\mbox{cos}\theta_{i}\mbox{cos}\varphi_{i}|=\frac{1}{2}$,}
the projection\emph{ }of the magnetic field $B_{z}>0$ with the porphyrin
rings and we study the resulting topological phases as a function
of the tilting angles $\theta_{i}$ ($\pm\varphi_{i}$ is fixed by
$\kappa$, $i=a,b$). Red regions are parameters for which values
of $\theta_{i}$ and $\varphi_{i}$ cannot yield $\kappa=\frac{1}{2}$,
and therefore, are not considered. (a) and (b) are diagrams for the
upper and lower energy exciton Hamiltonians $\mathcal{H}_{X}$ ($v=L,U$),
respectively. Light and dark blue regions denote topologically nontrivial
phases with Chern number $\mathcal{C}_{v}$ equal to -1 and 1, exhibiting
edge states with counterclockwise and clockwise exciton currents,
respectively. Switching the direction of the magnetic field to $B_{z}<0$
inverts these chiralities. Black lines correspond to topologically
trivial phases with $\mathcal{C}_{v}=0$ and are located along the
critical parameters $\theta_{b}=\pm\theta_{a},\pi\pm\theta_{a}$,
where the two sublattices become identical and the gaps of the respective
Hamiltonians vanish. \label{fig:phase_diagram}}
\end{figure}

A computational exploration of the parameter space reveals that $\mathcal{C}_{v}=\pm1$
for every tilting configuration that respects the specified constraints,
except for a measure zero set of parameters which yields $\mathcal{C}_{v}=0$,
when the two sublattices become identical at the critical values $\theta_{b}=\pm\theta_{a},\pi\pm\theta_{a}$,
and the gap between the two bands in each $\mathcal{H}_{v}$ closes.
Additionally, we can see that $\langle\boldsymbol{n}_{L}^{(i)}|J|\boldsymbol{m}_{L}^{(j)}\rangle=\langle\boldsymbol{n}_{U}^{(i)}|J|\boldsymbol{m}_{U}^{(j)}\rangle^{*}$,
so that every minimal loop that features a flux $\chi$ in $\mathcal{H}_{L}$
(see previous section), features the opposite flux $-\chi$ in $\mathcal{H}_{U}$,
implying that $\mathcal{C}_{L}=-\mathcal{C}_{U}$. Therefore, both
clockwise and counterclowise edge currents show up in every topologically
nontrivial configuration, except that they come at different energies
separated by $\sim2\Delta$. Yet, it is possible to control the energy
level ordering of these chiral currents by tuning the direction of
the magnetic field, as our numerical studies show that $\mathcal{C}_{L}=-\mbox{sgn}(B_{z})$.
This picture contrasts radically with that of the integer QHE in a
two-dimensional electron gas in the absence of a lattice, where the
direction of the magnetic field imposes a fixed direction of cyclotron
motion of the electrons and therefore, also the chirality of all the
edge currents. As an illustration of these ideas, Fig. \ref{fig:phase_diagram}
shows the topological phase diagram for the $\kappa=\frac{1}{2}$,
$B_{z}>0$ case. Given $\kappa$ and the fact that $\mbox{cos}\theta_{i},\mbox{cos}\varphi_{i}\in[-1,1]$,
we must restrict $\mbox{\ensuremath{\theta}}_{i}\in[-\mbox{acos}\kappa,\mbox{acos}\kappa]\cup[\pi-\mbox{acos}\kappa,\pi+\mbox{acos}\kappa]$.
The rest of the angles violate the condition of fixed $|\kappa_{i}|=\kappa$,
but a fraction of them still contains topologically nontrivial phases.
The characterization of this precise fraction is beyond the scope
of this article, but will be explored in the extension of this work.

Let us be more explicit by considering a particular point $(\theta_{a},\varphi_{a})=(-\frac{\pi}{3},0)$
and $(\theta_{b},\varphi_{b})=(0,\frac{\pi}{3})$ of this phase diagram
where $-\kappa_{a}=\kappa_{b}=\kappa=\frac{1}{2}$. We refer the reader
to Fig. \ref{fig:Eigenstates}, which is organized in a top and a
bottom panels (a) and (b), each of them containing three parts. We
show results for $\mathcal{H}_{L}$, with the conclusions for $\mathcal{H}_{U}$
being analogous except for opposite chirality of edge currents (energies
and dipoles are plotted in units of $J$ and $d$).\textbf{ }The upper
panel refers to the ideal case where the tilting angles of the porphyrins
are placed exactly at the mentioned values. The lower panel offers
a specific realization of disorder where each of the site angles has
been randomized with Gaussian noise with $0.13\pi$ standard deviation
about the ideal values. The left panels show the current density for
a particular eigenstate of $\mathcal{H}_{L}$ under OBC. These currents
are concentrated along the edges of the material, so they correspond
to exciton edge states and they flow clockwise, consistent with $\mathcal{C}_{L}=-1$.
Interestingly, in the disordered lattice, regardless of the tilting
randomization, the edge current and its chirality are still preserved.
In order to accentuate this effect, we add a potential barrier at
the left corner of the lattice, simulating an obstacle. The exciton
current simply circumvents the obstacle, keeping its delocalization
throughout, hence, exemplifying the properties of topological protection.
We have shown lattices with approximately 200 porphyrins, corresponding
to a reasonable number of molecules that remain coherently coupled
at cryogenic temperatures \cite{photoenergy}; this number might even
be a lower bound, as coherence size is limited by coupling to vibrations
and disorder, but the latter is somehow circumvented in these topological
systems. The center panels offer the energy diagrams of the respective
lattices under OBC along $\boldsymbol{y}$ and periodic boundary conditions
(PBC) along $\boldsymbol{x}$. For the ideal lattice, this corresponds
to two bulk bands as a function of quasimomentum $\boldsymbol{k}_{\boldsymbol{x}}$
together with edge states that span the gap between the latter from
$E\approx-2J$ to $2J$. The dispersion of the edge states is positive
and negative, corresponding to currents at the bottom and top edges
of the lattice. Note that these states of opposite dispersion merge
at $\boldsymbol{k_{x}}=0$ with the bulk bands. The analogous band
diagram is unavailable for the disordered lattice due to lack of translational
symmetry, so we simply collapse all the eigenenergies in the same
line. A study of the eigenstates reveals that the eigenstates between
$E\approx-0.8\, J$ and $0.4\, J$ exhibit mostly edge character.
We comment that, in fact, edge states seem to survive up to a large
amount of disorder, namely, with noise distributed at $\frac{\pi}{6}$
standard deviation. Finally, the right panels show the linear absorption
spectra of the lattice with OBCs along both directions $\boldsymbol{x}$
and $\boldsymbol{y}$. In analogy with J- and H- aggregates, most
of their oscillator strength is concentrated in relatively few bulk
eigenstates in the ideal lattice, although neither at top or bottom
of the bands, as opposed to the simple quasi-one-dimensional scenario
\cite{photoenergy}. This renders the edge states in the top panel
mostly dark, with the brightest edge state absorbing only 2.7\% of
the highest absorption peak in the spectrum. This fact is consistent
with the observation that, in the dipole approximation, only states
with $\boldsymbol{k}_{\boldsymbol{x}}=0$ are bright, but there are
no such states located at the edge in our particular model. Counterintuitively,
moderate amounts of disorder provide a solution to this problem, as
the edge states in this lattice borrow enough oscillator strength
from the original bulk states to yield peaks in the absorption spectrum
that are more experimentally accessible \cite{fidder:7880}, with
some edge absorption peaks attaining intensities of about 26\% of
the highest bulk bands. Hence, linear absorption spectra provide a
coarse signature of the edge states although no actual confirmation
of their topological character. In order to experimentally probe the
latter, we envision the use of near field optical microscopy, where
a metal tip locally creates excitons at the edge of the lattice and
spatially resolved fluorescence is used to detect the chirality of
the resulting exciton currents \cite{sunney}. This phenomenology
could also be inferred using far field microscopy, albeit at a coarser
spatial resolution \cite{akselrod}. The detailed proposal for the
experimental preparation and detection of the edge exciton currents
is a delicate subject on its own, and will be reported in an extension
of this work. Similarly, as noted at the beginning of the article,
we have omitted the discussion of interactions between excitons and
molecular vibrations, regarding them as weak compared to the coupling
$J$. Assuming a thermal scale $k_{B}T$ for the former, where $k_{B}$
is Boltzmann's constant and $T$ is the temperature, we require the
hierarchy of energy scales $k_{B}T\ll J\ll2\Delta$, which amounts
to cryogenic values of $T\approx4\,\mbox{K}$, given the values of
$J$ and $2\Delta$ in this article. We emphasize that the restriction
to cryogenic temperatures is not a fundamental one. It arises from
the modest values of Zeeman splitting $2\Delta$ attained at realistic
magnetic fields, and not from $J$, which can assume large values
upon appropriate chemical functionalization \cite{doi:10.1021/ja502765n}.
Hence, if $2\Delta$ can be enhanced otherwise (i.e., via optical
dressing), the domain validity of the model can be pushed to much
higher temperatures. For arbitrary exciton-vibrational couplings,
the current model needs to be adapted, and at present, it is not clear
under which general conditions will topologically non-trivial phases
survive, as vibrations might supress them if they serve as a thermal
dephasing bath, but may also sustain them nontrivially if they act
as a non-Markovian one.

\section*{Summary and Conclusions}

The present article introduces the concepts of topological phases
to the field of molecular excitonics. It does so by explicitly constructing
a topologically nontrivial model for Frenkel excitons in a two-dimensional
lattice of porphyrins. Important ingredients of the model are the
presence of two orbitally polarized excitons per porphyrin, the interaction
of the lattice with a perpendicular magnetic field, the anisotropy
of dipolar interactions between excitons, and the two-sublattice configuration
of tilted porphyrins, yielding two pairs of exciton energy bands.
The proposed system is a variant of the Haldane model, yielding one-way
exciton edge states that are robust against disorder, as we have shown
by calculations of topological invariants of the resulting energy
bands as well as by explicit simulations of finite lattices. An experimental
signature of these edge states is given by linear absorption spectra,
although the experimental confirmation of their topological character
requires more careful experiments, which will be proposed elsewhere.

We believe that our work is just one of many examples yet to be studied
of a new pool of strategies to engineer robust ``exciton wires''
that can efficiently transport light harvested energy. Among some
specific future directions, we plan to explore whether the coupling
of excitons with various spatially shaped electromagnetic fields such
as plasmons and optical cavities renders topologically nontrivial
motion of excitons. Furthermore, connections between this work and
theories of TRS breaking in quantum transport remain to be explored
\cite{zimboras,2014arXiv1405.6209L}. Another intriguing extension
of this work is the TRS version of the model, where no magnetic field
is present. Regarding the Q-band of each porphyrin as a pseudospin,
the anisotropy of the dipolar couplings may be regarded as a pseudospin-orbit
coupling, yielding excitonic TI analogues, or more precisely, to analogues
of topological crystalline insulators, which are a new class of materials
where orbital degrees of freedom together with spatial symmetries
of the lattice render topologically nontrivial band structures \cite{liangfu,hsiehfu}.
We hope to have convinced the reader that topological excitonics is
an exciting frontier of soft condensed matter and materials physics
research.

\begin{center}
\begin{figure*}
\begin{centering}
\includegraphics[scale=0.3]{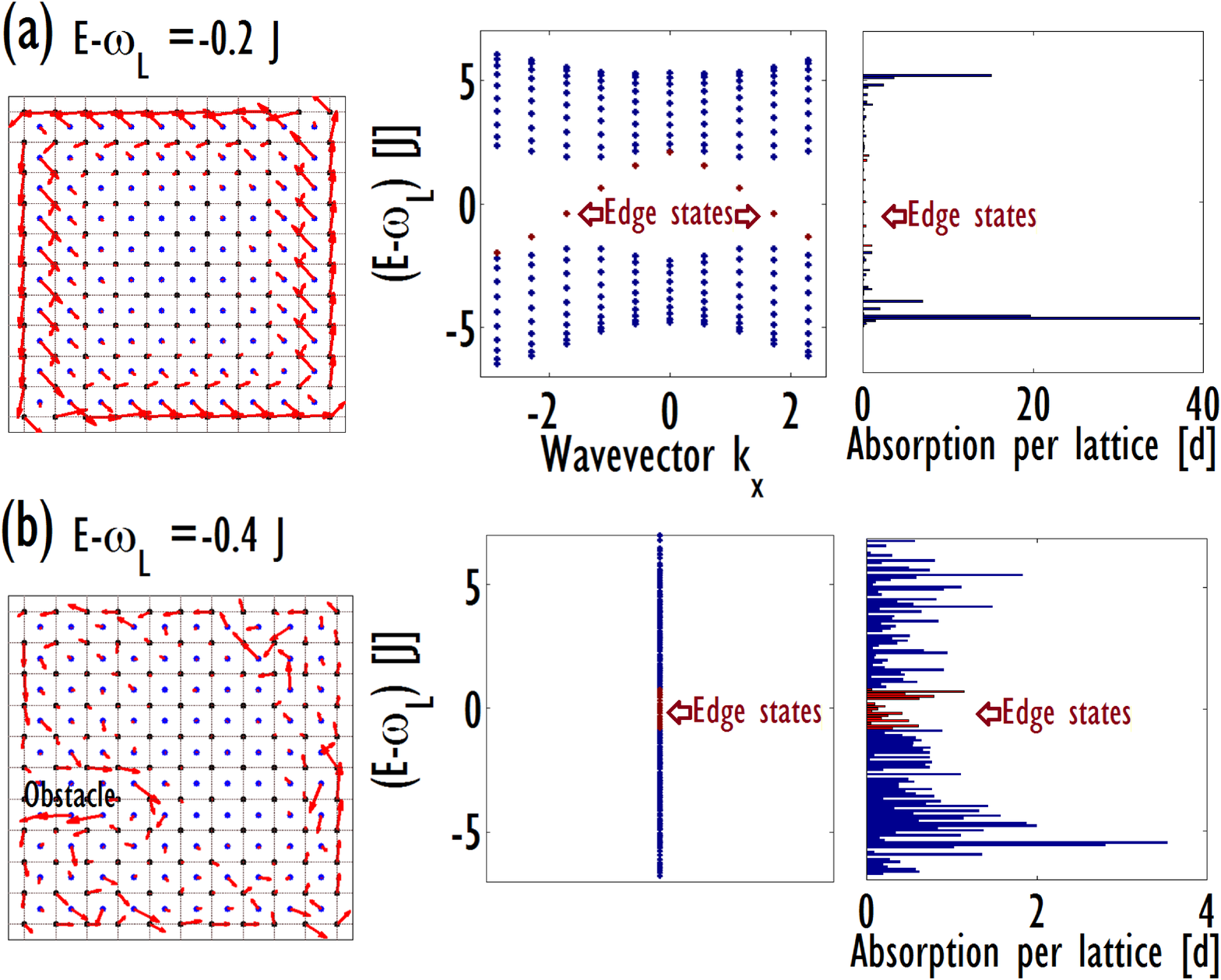}
\par\end{centering}

\protect\protect\caption{\emph{Eigenstates of the lower energy Hamiltonian }\emph{\noun{$\mathcal{H}_{L}$}}\emph{
for porphyrin tilting angles} $(\theta_{a},\varphi_{a})=(-\frac{\pi}{3},0)$
and $(\theta_{b},\varphi_{b})=(0,\frac{\pi}{3})$ \emph{and magnetic
field }$B_{z}>0$. These parameters yield a $\mathcal{C}_{L}=-1$
phase, which exhibits counterclockwise edge currents. (a)\emph{ }Ideal
lattice without disorder.\emph{ }The left panel shows the current
density for a particular edge state. The center panel depicts the
band diagram with bulk (blue) and edge (red) states. The positive
and negative dispersion for the edge states correspond to right and
left moving states which are localized at the lower and upper edges
of the sample. The right panel shows the absorption spectrum of the
lattice, indicating that most of the oscillator strength is primarily
concentrated in a few bulk states. (b) The analogous panels for a
disordered lattice, where the tilting angles for each porphyrin site
is randomized with noise distributed at $0.13\pi$ standard deviation.
Additionally, an obstacle (potential barrier) is added on the left
hand corner of the lattice. Note that the edge current density persists
with disorder, and in fact, circumvents the obstacle, remaining delocalized
across the edge. In the center panel, we show the corresponding density
of states, highlighting the region of energy where the states still
have a substantial edge character. The right panel shows the absorption
spectrum, where the disorder redistributes the oscillator strength
of the ideal lattice. The edge states borrow enough oscillator strength
to be measured experimentally in a linear absorption experiment. \label{fig:Eigenstates}}
\end{figure*}

\par\end{center}

\section*{Supplementary Information}

\emph{Topological characterization of the lattice Hamiltonian $\mathcal{H}_{L}$.---}
By imposing Periodic Boundary Conditions (PBC) along $\boldsymbol{x}$
and $\boldsymbol{y}$, $\mathcal{H}_{L}$ can be rewritten in quasimomentum
$\boldsymbol{k}$ space using the operators $a_{\boldsymbol{k}}^{\dagger}=\frac{1}{\sqrt{N_{x}N_{y}}}\sum_{\boldsymbol{n}}a_{\boldsymbol{n}}^{\dagger}e^{-i\boldsymbol{k}\cdot\boldsymbol{n}}$
and $b_{\boldsymbol{k}}^{\dagger}=\frac{1}{\sqrt{N_{x}N_{y}}}\sum_{\boldsymbol{n}}b_{\boldsymbol{n}}^{\dagger}e^{-i\boldsymbol{k}\cdot\boldsymbol{n}}$,
where $N_{x}$ and $N_{y}$ are the number of unit cells along the
$x$ and $y$ directions, and $\boldsymbol{k}=(k_{x},k_{y})$ can
take values in the Brillioun zone $-\pi\leq k_{x},k_{y}<\pi$ in discrete
steps of $\Delta k_{x}=\frac{2\pi}{N_{x}}$ and $\Delta k_{y}=\frac{2\pi}{N_{y}}$,
respectively. In terms of these operators, Eq. (\ref{eq:Hamiltonian_lattice})
becomes $\mathcal{H}_{L}=\sum_{\boldsymbol{k}}(\begin{array}{cc}
a_{\boldsymbol{k}}^{\dagger} & b_{\boldsymbol{k}}^{\dagger}\end{array})H(\boldsymbol{k})(\begin{array}{cc}
a_{\boldsymbol{k}} & b_{\boldsymbol{k}}\end{array})^{T}$, where $H(\boldsymbol{k})=\boldsymbol{d}(\boldsymbol{k})\cdot\boldsymbol{\sigma}+f(\boldsymbol{k})I$
describes a two-band model. Here, $\boldsymbol{\sigma}=(\sigma_{x},\sigma_{y},\sigma_{z})$
and $I$ are the vector of Pauli spin matrices and the two-by-two
identity matrix, and $\boldsymbol{d}(\boldsymbol{k})=(d_{x}(\boldsymbol{k}),d_{y}(\boldsymbol{k}),d_{z}(\boldsymbol{k}))$
and $f(\boldsymbol{k})$ are a $\boldsymbol{k}$ dependent vector
and scalar given by,

\begin{subequations}

\begin{eqnarray}
d_{x}(\boldsymbol{k}) & = & A_{1}\mbox{cos}\Bigg(\frac{k_{x}+k_{y}}{2}\Bigg)+A_{2}\mbox{cos}\Bigg(\frac{k_{x}-k_{y}}{2}\Bigg),\label{eq:dx}\\
d_{y}(\boldsymbol{k}) & = & B_{1}\mbox{cos}\Bigg(\frac{k_{x}+k_{y}}{2}\Bigg)+B_{2}\mbox{cos}\Bigg(\frac{k_{x}-k_{y}}{2}\Bigg),\label{eq:dy}\\
d_{z}(\boldsymbol{k}) & = & C_{1}\mbox{cos}(k_{x})+C_{2}\mbox{cos}(k_{y}),\label{eq:dz}\\
f(\boldsymbol{k}) & = & D_{1}\mbox{cos}(k_{x})+D_{2}\mbox{cos}(k_{y}),\label{eq:f(k)}
\end{eqnarray}

\end{subequations}where $A_{1}=2\Re J_{ab,NE}$, $A_{2}=2\Re J_{ab,NW}$,
$B{}_{1}=-2\Im J_{ab,NE}$, $B_{2}=-2\Im J_{ab,NW}$, $C_{1}=J_{aa,E}-J_{bb,E}$,
$C_{2}=J_{aa,N}-J_{bb,N}$, $D_{1}=J_{aa,E}+J_{bb,E}$, and $D_{2}=J_{aa,N}+J_{bb,N}$
(see main text for the definitions of these coupling terms). Using
this parametrization, one can readily compute the Chern number $\mathcal{C}_{L}$,
a topological invariant that, due to the bulk-edge correspondence
\cite{bernevigbook}, yields the (integer) number of edge states per
quasimomentum under Open Boundary Conditions (OBC) lying in the gap
of the spectrum of $\mathcal{H}_{L}$ or alternatively $H(\boldsymbol{k})$,
whose energy bands are given by,

\begin{equation}
\epsilon(\boldsymbol{k})=f(\boldsymbol{k})\pm|d(\boldsymbol{k})|.\label{eq:energy_bands}
\end{equation}
Defining the unit vector $\hat{d}(\boldsymbol{k})=\frac{\boldsymbol{d}(\boldsymbol{k})}{|\boldsymbol{d}(\boldsymbol{k})|}$,
$\mathcal{C}_{L}$ can be calculated as a integral in $\boldsymbol{k}$
space \cite{tknn,bernevigbook},

\begin{equation}
\mathcal{C}_{L}=\int_{-\pi}^{\pi}dk_{x}\int_{-\pi}^{\pi}dk_{y}\mathcal{B}(k_{x},k_{y}).\label{eq:chern}
\end{equation}
where the Berry curvature in terms of $\hat{d}(\boldsymbol{k})$ is
given by $\mathcal{B}=\frac{1}{4\pi}\hat{d}\cdot\left(\partial_{k_{x}}\hat{d}\times\partial_{k_{y}}\hat{d}\right)$.
The simplicity of the parametrization in Eqs. (\ref{eq:dx})--(\ref{eq:f(k)})
affords an analytical expression for $\mathcal{C}_{L}$,

\begin{widetext}

\begin{eqnarray}
\mathcal{C}_{L} & = & \frac{\mbox{Numerator}}{\mbox{Denominator}},\nonumber \\
\mbox{Numerator} & = & (A_{2}B_{1}-A_{1}B_{2})[C_{1}\mbox{cos}(2k_{x})+(C_{1}-C_{2})(-3+2\mbox{cos}(k_{x})\mbox{cos}(k_{y}))-C_{2}\mbox{cos}(2k_{y})],\nonumber \\
\mbox{Denominator} & = & 32\pi\{[C_{1}\mbox{cos}(k_{x})+C_{2}\mbox{cos}(k_{y})]^{2}+\frac{1}{2}[(A_{2}^{2}+B_{2}^{2})(1+\mbox{cos}(k_{x}-k_{y}))\nonumber \\
 &  & +2(A_{1}A_{2}+B_{1}B_{2})(\mbox{cos}(k_{x})+\mbox{cos}(k_{y}))+(A_{1}^{2}+B_{1}^{2})(1+\mbox{cos}(k_{x}+k_{y}))]\}^{3/2}.\label{eq:denominator}
\end{eqnarray}

\end{widetext} The sign of $\mathcal{C}_{L}$ indicates the chirality
of the edge state and topologically nontrivial phases are characterized
by a nonvanishing $C_{L}$.

A word of caution follows about the correct interpretation of $\mathcal{C}_{L}$
in the context of this exciton problem, besides the one already given.
First, were $\mathcal{H}_{L}$ to describe an electronic rather than
an exciton problem, not every band structure given by Eq. (\ref{eq:energy_bands})
satisfies the condition of an electronic insulator: due to the $\boldsymbol{k}$-dependent
offset $f(\boldsymbol{k})$, there is not always an gap where one
can place the Fermi energy such that it does not cross any of the
bulk bands. This issue deteriorates topological protection allowing
scattering into the bulk. However, for the cases that qualify as an
insulator, such as the example described in the main text, $\mathcal{H}_{L}$
represents a Chern insulator which exhibits a quantized transverse
conductance $\frac{e^{2}}{h}\mathcal{C}_{L}$ ($e$ is charge of an
electron, and $h$ is Planck's constant) under a weak voltage bias,
giving an experimental interpretation to the meaning of $\mathcal{C}_{L}$
\cite{tknn}. Our case is different as we are limiting ourselves to
single excitation effects, rather than considering a macroscopic occupation
of the states, as in an actual electornic insulator, and the occupation
of the bands occurs through light and not through a difference in
electrochemical potential. Yet, it is an intriguing problem to find
an excitonic observable that directly corresponds to $\mathcal{C}_{L}$.
One could imagine designing a protocol that occupies the edge states
and not the bulk bands, and which applies a bias potential to the
excitons via, say, mechanical strain \cite{strain}. The measurement
of an exciton current in this situation would be related to $\mathcal{C}_{L}$.
As noted in the main text, a more careful description of such an experimental
protocol involving near field spectroscopic techniques will be defined
in the extension of this work, so for now, we will content ourselves
with the first interpretation of $\mathcal{C}_{L}$ in terms of the
number of edge states.

Even though everything in this section referred to $\mathcal{H}_{L}$,
the analogous conclusions apply to $\mathcal{H}_{U}$. In fact, one
can easily show that Eq. (\ref{eq:denominator}) is also valid for
$C_{U}$ provided the corresponding dipolar couplings are used.

\emph{Symmetries of $\mathcal{C}_{v}$.---} In the text, we argued
that $\mathcal{C}_{L}=-\mathcal{C}_{U}$. This claim can be restated
as $C_{v}$ changing sign upon switching the $+$ and $-$ labels,
$\{|Q_{+}^{(i)}\rangle,|Q_{-}^{(i)}\rangle\}\leftrightarrow\{|Q_{-}^{(i)}\rangle,|Q_{+}^{(i)}\rangle\}$.
By keeping track of the various matrix elements, we can see that $C_{v}$
also reverses sign upon the transformation $\kappa_{i}\leftrightarrow-\kappa_{i}$
across both sublattices. However, inverting the projection of the
magnetic field on the porphyrins, $\kappa_{i}\leftrightarrow-\kappa_{i}$,
also reassigns the upper and lower energy states $|Q_{L}^{(i)}\rangle$
and $|Q_{U}^{(i)}\rangle$ between $|Q_{+}^{(i)}\rangle$ and $|Q_{-}^{(i)}\rangle$,
switching $+$ and $-$ labels. Hence, the value of $\mathcal{C}_{L}$
(and consequently, of $\mathcal{C}_{U}$) is fixed across tilting
configurations, and by computation seen to correspond to $\mathcal{C}_{L}=-1$
for $B_{z}>0$. Since $B_{z}$ also switches $+$ and $-$, $\mathcal{C}_{L}=1$
for $B_{z}<0$.

 \bibliographystyle{unsrt}

\section*{Acknowledgements}

J.Y.Z. is grateful to Dr. I. Kassal for introducing him to the topic
of topological phases, to Prof. B. Halperin and Dr. X. Andrade for
discussions, and to Prof. O. Starykh for kindly sharing his notes
on the subject. All the authors would like to thank Prof. C. Laumann
for discussions at the early stages of the project. J.Y.Z. and A.A.G.
are supported by an Energy Frontier Research Center funded by the
US Department of Energy, Office of Science, Office of Basic Energy
Sciences under Award Number DESC0001088. N.Y. acknowledges support
from DOE (FG02-97ER25308). Finally, S.K.S. and A.A.-G. are supported
by the Defense Threat Reduction Agency grant HDTRA1-10-1-0046.
\end{document}